# NetO-App: A Network Orchestration Application for Centralized Network Management in Small Business Networks

Dewang Gedia[1] and Levi Perigo[2]

[1] Interdisciplinary Telecom Program, University of Colorado Boulder
530 UCB, Boulder, Colorado, USA 80309
Dewang.Gedia@colorado.edu, Levi.Perigo@colorado.edu

## Abstract

*Software-defined networking (SDN) is reshaping the networking paradigm. Previous research shows that SDN has advantages over traditional networks because it separates the control and data plane, leading to greater flexibility through network automation and programmability. Small business networks require flexibility, like service provider networks, to scale, deploy, and self-heal network infrastructure that comprises of cloud operating systems, virtual machines, containers, vendor networking equipment, and virtual network functions (VNFs); however, as SDN evolves in industry, there has been limited research to develop an SDN architecture to fulfil the requirements of small business networks. This research proposes a network architecture that can abstract, orchestrate, and scale configurations based on small business network requirements. Our results show that the proposed architecture provides enhanced network management and operations when combined with the network orchestration application (NetO-App) developed in this research. The NetO-App orchestrates network policies, automates configuration changes, and manages internal and external communication between the campus networking infrastructure.*

## 1. Introduction

Software-defined networking (SDN) pronounced its presence in networking, but the attempts to create SDN architectures that can address the needs of small business networks have been limited [13]. According to Cisco Systems, small business networks require highly secure and reliable data networks that meet rigorous requirements such as remote workers, accessing customer data from any place and time, and cost-effective support of new applications [31]. Small business network infrastructures that employ network elements such as cloud operating systems, virtual machines (VMs), containers, traditional vendor networking equipment, and virtual network functions (VNFs); constantly need efficient control and configuration management mechanisms to dynamically cater to changing workloads. Such an environment is subjected to software and hardware restrictions, repetitive deployments and configurations, and dynamic business requirements. Small business enterprises, which are typically defined as possessing less than 500 employees in the United States and less than 250 employees in Europe, need to adopt an infrastructure that is efficient to configure and manage, inexpensive to deploy and operate, highly scalable, easy to operate, and secured from internal and external threats [32,17].

While SDN can be difficult to define, the Open Networking Foundation (ONF) defines an SDN architecture as a networking model that is directly and programmatically configured, decouples the network control functions from the forwarding functions, logically centralizes the control, and is open standards-based and vendor-neutral [3]. In a small business networking environment, the infrastructure incorporates both SDN and traditional devices and must use an architecture that can flexibly manage both traditional and SDN domains [13].

Traditional network engineering relies on device configuration via the command line interface (CLI) and does not scale to meet the complexity of multi-vendor SDN/traditional networks in small businesses. Programmability of traditional devices is cumbersome because they lack open, programmable interfaces, which prohibits developers from programming the network in the most efficient method [21,22]. Furthermore, integrating SDN and traditional networks is difficult due to the disparities between how they function: traditional networks operate with the help of MAC address tables and routing tables, whereas SDN with OpenFlow uses flow entries in flow tables. These disparities need a different methodology to integrate as a system, and research indicates that only a limited number of tools can handle these problems efficiently [23].

Network automation reduces the manual effort required for completing routine tasks and decreases the amount of human error caused by traditional, manual CLI configurations. Starting with scripting and progressing to intelligent network control and efficient translation and deployment of network plans and policies, network automation is a key tool to facilitate traditional network management and operations. While using information from configuration files and deploying routine configurations onto multiple network devices is a step towards automation, this approach can be made more dynamic by creating a graphical user interface (GUI) that automates configuration from minimal user input, simplifies the process, abstracts the network infrastructure from the programmer, because it does not require the programmer to know vendor-specific CLI commands, and reduces the number of misconfigurations [25].

Network programmability coupled with network automation can address SDN and traditional network limitations and can also provide a better platform for centralized configuration management of the cloud infrastructure in small business networks. Cloud computing is a rapidly growing paradigm for consuming data center resources in the form of Services: Platform-as-a-Service (PaaS), Infrastructure-as-a-Service (IaaS), and Software-as-a-Service (SaaS) [15]. Private cloud offers small business networks control over the infrastructure, choice of hardware/software tools, and offers control over the desired network security; thus, minimizing the scope of network vulnerability from external attacks. The benefits network virtualization and cloud computing offer when combined with SDN and network automation provide a framework that is suitable for small businesses.

The remainder of the paper is organized as follows: Section 2 provides a review of the existing body of knowledge, state of the art applications, and how our scheme extends it. Sections 3 and 4 describe the methodology and results of our experiment respectively. Section 5 concludes our research and addresses scope for future enhancements.

## 2. RELATED WORK

SDN has changed management of network infrastructure by decoupling the network control plane and the data plane [17]. With the help of ONF, there has been wide scale industry adoption of the OpenFlow protocol as the standard southbound interface (SBI) to communicate with pure and hybrid OpenFlow SDN switches. Although there have been attempts to create network architectures that are easily manageable, scalable, fault-tolerant, and inexpensive, there has been limited results that meet all these requirements for small businesses [13].

Small business network environments are constrained by limited software, hardware, and network capabilities. They also are hindered by repetitive tasks, limited administrative skillset, and the capability to dynamically adapt to constantly changing user workloads. Furthermore, the confined financial budgets dedicated to small business network environments prove to be a monumental restriction. In such an agile and restrictive environment, it becomes essential for the small business network infrastructure to meet or exceed these minimum requirements to have state-of-art network facilities with limited resources. The NetO-App developed from this study addresses these limitations by using free and open source software, a user-friendly web interface, and provides an advantage of having an automated and orchestrated infrastructure that reduces operating cost and increased ROI [27]. NetO-App efficiently addresses these requirements by using OpenStack Kolla as an orchestrator that deploys Docker containers that are lightweight, quickly scalable, require less storage space thus, catering to small business network requirements.

With the advent of cloud computing, more Internet applications such as DNS, DHCP, and web servers are deployed in the cloud. To manage these applications, a greater level of automation and orchestration is required. SDN helps build a level of abstraction and orchestration for VM management where hypervisors leverage the real-time network information before migration to minimize network-wide communication costs of resulting traffic dynamics [1]. A large-scale SDN capable infrastructure, the OF@TEIN playground, was initially targeted to build and operate OpenFlow enabled networks, but shifted its efforts to establish an open and shared consortium for new potential collaborators with the intention to build and operate a federated multi-site SDN-Cloud-leveraged infrastructure using the ONOS SDN controller, OpenStack cloud, and Quagga router [2]. Using Quagga to facilitate the transition from a traditional network to an SDN has provided a platform for exchanging border gateway protocol (BGP) routing information [3]. Although such an architecture provides inter-platform networking capabilities, it still lacks centralized application for configuration and management of a multi-platform infrastructure that our proposed model provides.

One of the motivations for developing SDN was to overcome challenges faced by data centers. An architectural framework provided by ONF is Central Office Re-architected as Data Center (CORD). This platform helps service providers deliver a cloud-native, open, and programmable platform to enable services to end-users [4]. This architecture primarily enables residential, mobile, and enterprise subscribers to appropriately route traffic using defined network policies residing on the XOS (CORD controller) node. While CORD tackles policy based routing through XOS, the framework, lacks necessary components for deploying and scaling VMs/containers, and incorporating multi-vendor traditional network hardware present in small business network environments. The proposed NetO-App specifically addresses such business network requirements and appropriately automates the VM/container deployment.

Open Network Automation Platform (ONAP) aims to provide a comprehensive platform for the real-time deployment and policy-driven orchestration of network functions for cloud providers and operators to automate new services [20]. Additionally, it offers the capability to monitor the service behavior based on the specified design and provides healing capabilities by scaling the resources to adjust any demand variations. Although the ONAP platform can deliver service design, creation, and lifecycle management in an OpenStack VM environment, it lacks capabilities to host and monitor the VNFs in a container platform which increases the expense of this platform due to the exponential storage space required for VMs over containers. Furthermore, the amount of dedicated hardware required to operate ONAP is difficult for small business network environments. NetO-App addresses these limitations by leveraging Ansible to proactively monitor containers and VMs hosted in network, and can be deployed on a single server. Another disadvantage of ONAP is that it has a complex architecture and needs a thorough understanding of every module to tailor desired services needed for a small business network. NetO-App provides a simple architecture that is easy to control through a user-friendly web portal to dynamically create VMs/containers.

The COSIGN project highlighted the integration of SDN controllers and the OpenStack orchestrator for optimizing the selection of resources in a virtual data center [5]. A fabric topology using SDN helps overcome the bandwidth utilization and network scalability challenges posed by fat tree topology [6]. However, the approaches still fail to deliver self-healing incident-response (pre-defined) capabilities in an orchestrated cloud environment that NetO-App addresses.

As shown in [7], integrating OpenStack with SDN provides benefits when managing complex and virtualized applications. With a better GUI, it is easy to manage SDN topologies which were demonstrated by integrating OpenStack and the Ryu SDN controller [7]. While transitioning from traditional networks toward centralized SDN, SDN placement planning can help achieve better controllability in the early 70% of the deployment [8]. Using the OpenDaylight (ODL) SDN controller for achieving network programmability (flow control and network isolation) in an OpenStack environment through the provided Neutron plugin can help provide centralized management in cloud operating systems as well [9]. However, the Neutron service fails to provide the necessary encapsulation for the traffic flowing between different tenants. Our proposed model overcomes this limitation by employing the OpenContrail SDN controller.

As new SDN design architectures emerge, a framework is required to manage and coordinate different implementations. The concept of a network hypervisor was introduced in [10] which provides a platform to use the existing low-level application programmable interfaces (API) provided by different SDN implementations in an autonomous system and convert it to high-level APIs. This can ease the task of creating an SDN; however, it fails to leverage capabilities to remotely configure and manage infrastructure through a centralized application. It was found in [12] that when integrated into an OpenStack environment, ODL has inferior performance in terms of delay and throughput when compared to Floodlight and Ryu, but the ODL controller showed higher resiliency.

In this paper, the primary research question we answered was "**Can an application provide centralized network management to configure and manage multi-platform, software-defined, and traditional networking environments for small business networks?**" This research question was strategically divided into subproblems that addressed an individual technological research aspect to collectively answer the primary research question.

A. Can we achieve orchestrated control to configure and manage multi-platform network elements using an application?
B. Can we achieve a scalable and resilient control infrastructure for remote network management and configuration?
C. Can BGP be used to facilitate interconnectivity between VM/containers, SDN, and traditional networks?

The contribution of this paper is to design and implement a network architecture and application that can be used together to orchestrate multi-platform environments, such as virtualized cloud, SDN, and traditional networks in small business networks.

## 3. RESEARCH OVERVIEW

### 3.1. Environment

The network architecture developed for this research was comprised of an overlay and underlay network (Fig. 1). The overlay network consists of a multi-node OpenStack setup which is operating on five x86 servers, VMs with Docker containers for specific applications, OpenContrail and Floodlight SDN controllers, and the network orchestration application (NetO-App) developed from this study. The underlay network is composed of x86 servers running Ubuntu, OpenFlow-capable switches including Arista, Cisco, Dell, HP, OvS, OpenSwitch, and Pica8, which establish an OpenFlow v1.3 channel to the Floodlight container for the SDN, and traditional networking equipment including ADTRAN, Arista, Cisco, and Juniper.

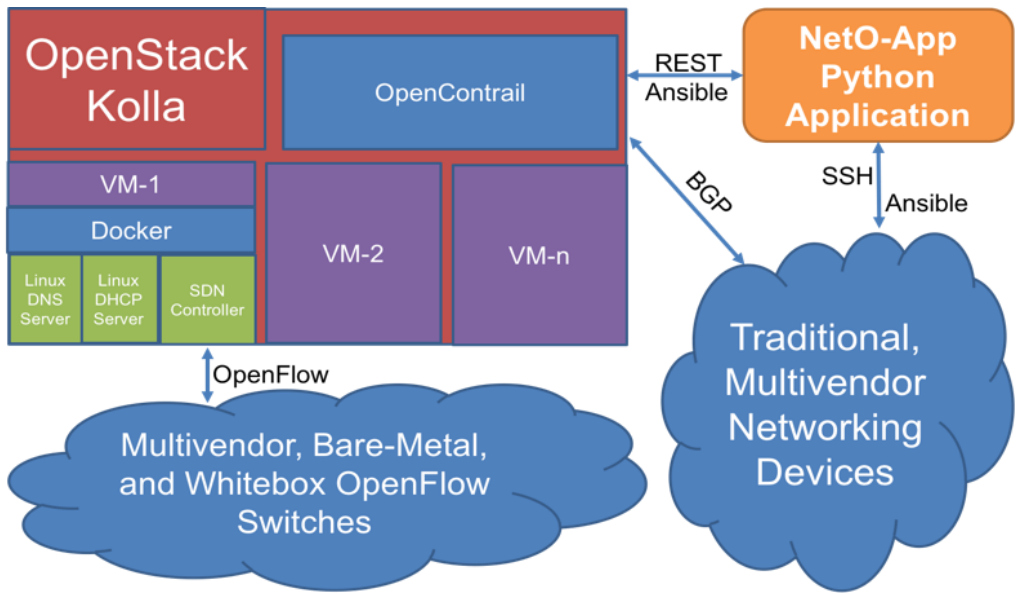

Figure 1: Network Architecture Design

### 3.2. Hypervisor and Containers

For this study, OpenStack was selected as the private-cloud OS because it provides a virtualized, cloud infrastructure that leverages abstraction, orchestration, and automation capabilities for the network infrastructure. In the network architecture created as a result of this research, we implement a specific project forked from the master OpenStack project, OpenStack Kolla. OpenStack Kolla provides Representational State Transfer (REST) APIs, greater programmability, increased resource management, network virtualization, visibility and real-time monitoring, as well as multi-tenancy support. OpenStack Kolla deploys the OpenStack services in containers; thus, reducing the underlying server storage and achieves rapid boot time of the services with auto-scaling functionality [15]. The auto-scaling functionality of OpenStack Kolla implements a service called Senlin which facilitates automatic VM redundancy because it can dynamically distribute dedicated compute resources due to failure, user-defined thresholds, and utilization.

Containers have proved to be advantageous over their VM counterpart because of their portability, highly responsive lifecycle management, orchestration, agility, and elasticity [11]. In this research, we have selected Docker containers to be used within VMs in the OpenStack Kolla environment. The services we deployed in Docker were the DHCP server, DNS server, and the SDN Floodlight controller.

### 3.3. SDN Controllers

SDN implementations are evolving, but there has been limited research on providing seamless interconnectivity between varied platforms [26]. There is a need to adopt a routing mechanism for inter-connectivity between VMs/containers and both SDN and traditional networking devices. To achieve this, our network architecture implements an OpenContrail SDN controller to provide networking service for the virtualized OpenStack Kolla environment, the Floodlight controller for the OpenFlow SDN, and the vendor routers for the traditional network. This is beneficial in this research because OpenContrail has an intuitive Python REST API for automation and utilizes BGP to connect both SDN and traditional networks; thus, serving as an optimal solution for bringing inter-platform connectivity between the OpenStack Kolla environment and both the traditional and SDN devices.

The Floodlight SDN controller was implemented in this research to control the SDN infrastructure via OpenFlow. Floodlight was selected as the SDN controller because it is well-tested and has well-defined APIs. The APIs of Floodlight provide a documented menu which allows researchers to create network control and management applications with relative ease [12]. This was critical in this research because of the need for abstraction and orchestration between platforms.

### 3.4. The NetO-App

The NetO-App developed from this research to manage and configure multi-platform environments in small business networks was built using the Python programming language. Python was selected because it has an object-oriented design that provides high-level, built-in data types, user-friendly data structures, and support libraries [14]. The NetO-App is comprised of two primary modules: the abstraction module and the implementation module (Fig. 2).

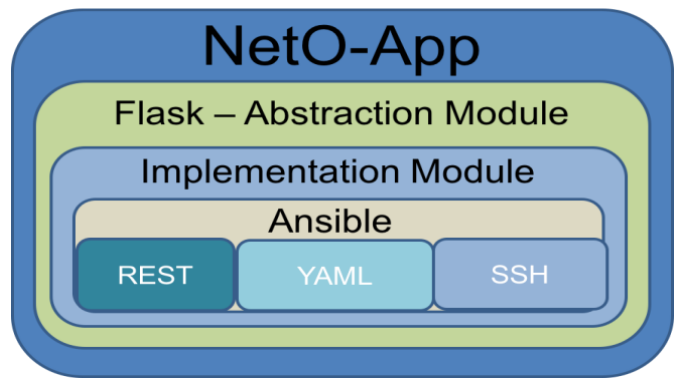

Figure 2: The Python NetO-App model

### 3.5. The abstraction Python module

The abstraction module provides a front-end interface for configuration, management, and network automation using the Python Flask web framework. This module provides a user-friendly GUI that abstracts the multi-platform network infrastructure to the user and provides the user with flexibility to select functions to orchestrate the entire network architecture.

### 3.6. The implementation Python module

The implementation module is comprised of the following sub-modules: Ansible, YAML, REST, and SSH. Each of the sub-modules is used to configure and manage various parts of the proposed network infrastructure environment.

To interact with network nodes for configuration, the implementation module of NetO-App uses the Python-based infrastructure automation framework Ansible. Ansible is a configuration and automation tool [18] that is prevalent in the industry for managing and configuring network devices and servers [19]. It was designed to make remote configurations quicker using SSH [18]. Ansible is agentless, so it does not require an agent to be installed on the client; instead, it uses SSH to push changes to the remote server or host defined in Playbooks. Playbooks describe hosts and tasks and are defined in YAML Ain't Markup Language (YAML) format [19]. Apart from SSH, Ansible can also communicate using APIs; thus, extending the number of network elements that can be configured using Ansible. For this research, Ansible is a tool that can be used to configure the cloud, SDN, and traditional networks.

### 4. RESULT AND ANALYSIS

In the experiments, we answered the sub-problems defined in section II to ultimately answer the primary research question.

## 4.1. Research sub-problem: Can we achieve orchestrated control to configure and manage multi-platform network elements using an application?

This research sub-problem guided the creation of NetO-App. The NetO-App combined with the proposed network infrastructure from this study provides a solution for centralized network management of cloud, SDN, and traditional networks that can dynamically push network configurations to the overlay and underlay network devices using SSH, Ansible, and REST. To make the solution user-friendly, we have developed a Flask front-end that abstracts the specific commands from the user. The user can make changes via the GUI, and then the back-end implementation module of NetO-App will execute the appropriate platform configuration scripts such as Ansible, REST, or SSH. This provides convenience to the user, without having to memorize multiple vendor specific commands, or understand a programming language. This abstraction layer is important to the research because small business network environments do not have the dedicated, skilled resources that other institutions have; thus, an intuitive web front-end is paramount to the success of the small business network model. The NetO-App communicates via REST and Ansible/SSH to dynamically configure and manage VMs within the OpenStack Kolla environment and uses Ansible/SSH to update or change configurations as described within Ansible Playbooks. When the user clicks on desired functions in the GUI, an appropriate Python script is executed, which invokes the Paramiko module inside of Ansible to SSH into either the VM, software-defined device, traditional device, or all. In the event of making changes to an existing configuration on any platform, the application checks for the present configuration and only pushes configurations that need to be updated; thus, providing efficiency and consistency across relevant nodes.

The NetO-App takes the user input based on mandatory parameters - hostname of node (on which VMs/containers would reside), tenant name, number of VMs/containers to deploy (preconfigured with desired packages), and the type of VMs/containers (Ryu controller/ONOS controller/ODL controller/Mininet/OVS). Optional parameters include – validation checks (verify VMs/containers are configured per the user-defined requirements), and fresh install (revert the VMs/containers back to the clean state). Once the user has defined the parameters, NetO-App specifically uses Ansible/SSH to instantiate and configure VMs/containers per the requirement.

The NetO-App is beneficial in small business networks because it makes it easier for the network user to orchestrate, manage, and configure a multi-platform environment consisting of cloud, SDN, and traditional networks from a centralized point without having to understand the underlying vendor-specific CLIs, programming language, or cloud operating systems. Furthermore, the NetO-App is free and open source software which appeals to small business networks' budgetary constraints.

## 4.2. Research sub-problem: Can we achieve a scalable and resilient control infrastructure for remote network management and configuration?

In our proposed model, we deployed multiple services, such as Floodlight SDN controllers within Docker containers on the OpenStack Kolla environment to provide a scalable and resilient control infrastructure to manage and configure the SDN. We could provide resiliency in the event of the failure of a VM because the Senlin service that is enabled within the OpenStack Kolla environment constantly monitors VMs for utilization. For example, when the threshold compute value of the containerized SDN controller is reached, the Senlin service spawns the secondary controller container and disables the primary controller, providing high availability of the control plane and optimizes resource efficiency. Furthermore, the OpenStack Kolla environment is hosted on multiple physical servers, which provides dynamic resilience and physical redundancy for the virtualized control platform.

To understand the amount of downtime achieved through this automated OpenStack scaling approach, we conducted a test that compared a manual configuration to an automated configuration. It was found that manually it took 57 seconds to create a secondary SDN controller whereas, Senlin could detect and configure a secondary SDN controller (container) in 1.2 seconds, and do it automatically without manual intervention. We incrementally added the configuration time based on the number of devices to understand the average number of configured containers in a stipulated time. Fig. 3 demonstrates that it took approximately 15 seconds to configure 10 controllers. This particularly demonstrates that the current OpenStack approach deployed in this research can reduce the downtime to seconds instead of minutes, and is fully automated requiring no manual intervention to self-heal.

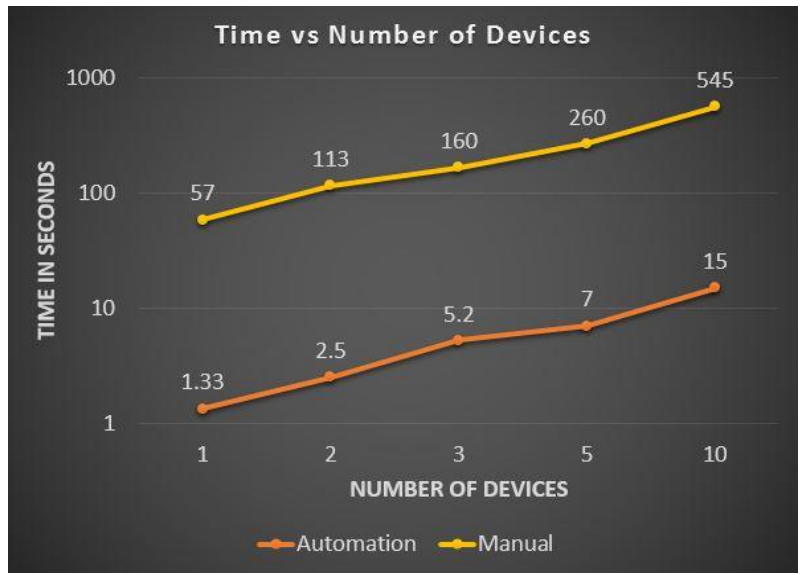

Figure 3: Time taken to configure SDN controller manually versus OpenStack automated approach

Therefore, to achieve a scalable and resilient control infrastructure for network management and configuration the research from this study answered the sub-problem by designing an architecture consisting of a virtualized cloud environment using OpenStack, Docker containers for required, lightweight service functionality, and the NetO-App which monitors and manages this environment to dynamically resolve physical and virtual failures and configuration quickly.

### 4.3. Research sub-problem: Can BGP be used to facilitate interconnectivity between VM/containers, SDN, and traditional networks?

To provide the virtual and physical L2 and L3 external connectivity, we used OpenContrail as a networking solution that works with the neutron service of OpenStack to provide cloud networking capabilities [24]. The OpenContrail architecture consists of two main components: vRouter and Controller. The OpenContrail controller uses Extensible Messaging and Presence Protocol (XMPP) to communicate with the vRouters and BGP/NETCONF to communicate with the traditional networking devices. As shown in Fig. 4, the control node present within the OpenContrail controller is responsible for processing routing information and applying them to the forwarding table of the vRouter service that handles networking for the OpenStack Kolla environment. To exchange routing information with the traditional vendor routers, the control node uses BGP. Thus, the use of BGP by the control node helps provide connectivity between VMs/containers and SDN devices with the traditional BGP speaking vendor routers.

OpenContrail provides REST APIs that are used by NetO-App to dynamically orchestrate the configuration of the VNFs inside of OpenStack. Therefore, the research answers this sub-problem by using OpenContrail as the VNF manager within OpenStack to communicate between the virtualized SDN and the traditional networking environment via BGP. Additionally, OpenContrail allows NetO-App to centralize control of this platform through the built-in OpenContrail REST API.

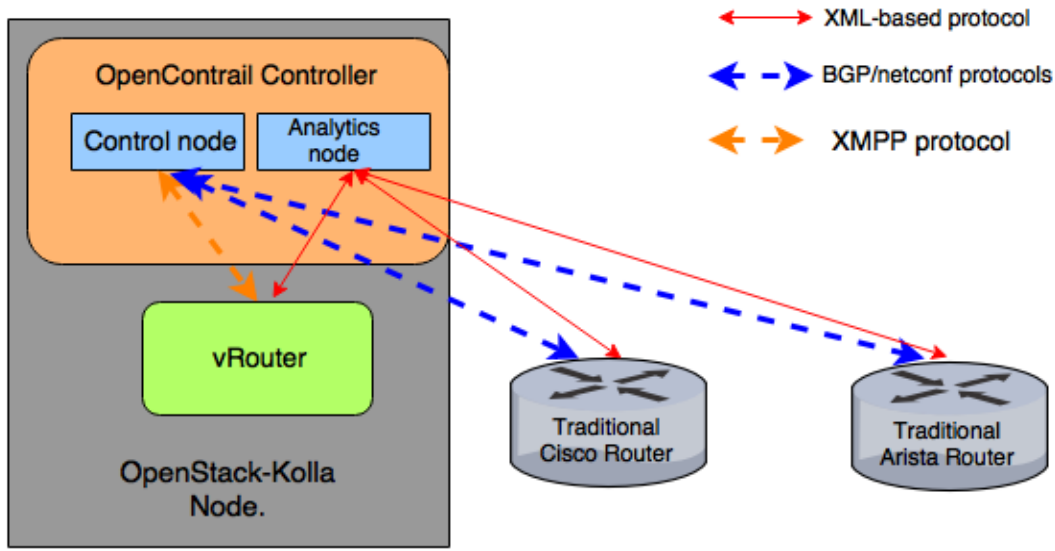

Figure 4: OpenStack-OpenContrail Networking Diagram

## 5. CONCLUSIONS

In this paper, we propose a network architecture and centralized network management application that can configure and manage multi-platform, software-defined, and traditional networking environments for small business networks. The small business network environment is subjected to limited software/hardware resources, repetitive deployments and configuration changes, dynamic adaptability to changing business requirements on a limited budget. It is essential for a small business organization to adopt an automated and orchestrated infrastructure that is efficient to configure and manage, inexpensive to deploy and operate, highly scalable, and provides ease of operation. We achieved this by designing an environment consisting of SDN, traditional networks, and cloud architectures. By utilizing OpenStack Kolla, we created a scalable and resilient control infrastructure for remote network management, configuration, scalability, and resiliency. OpenContrail facilitated interconnectivity between VMs, SDN, and traditional networks, through BGP and the NetO-App developed from this research implemented a user-friendly Flask front-end to abstract the underlying technologies from the user by utilizing Ansible, SSH, and REST to orchestrate the multi-platform cloud, SDN, and traditional network. Small business institutions can deploy the network architecture and NetO-App designed from this research to create a low-budget optimized network that is platform independent, centrally managed, resilient, scalable, easy to use, and inexpensive to implement.

Currently, this research focused on small business networks, but a future scope could enhance this by targeting other sectors with similar requirements, such as academia. Furthermore, NetO-App has minimal self-healing capabilities. NetO-App is able to execute predefined tasks upon a failure of a respective service. The future scope of this research can be improved by including a module for greater self-healing capabilities using TensorFlow (machine learning) for various failure scenarios [30]. This can be achieved by performing big data analysis and dissecting traffic parameters and performing actions to correct errors through TensorFlow. Specifically, Grafana [28] and Platform for Network Data Analytics (PNDA) [29] provide a platform for carrying out such analysis which serves as a future scope of the research.

## AUTHORS

**Dewang Gedia, University of Colorado Boulder**

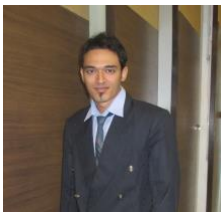

Dewang Gedia is a Ph.D. student at the University of Colorado Boulder having primary research focus in Network Functions Virtualization and Software Defined Networks domain. He achieved his Master's degree from Interdisciplinary Telecom Program (ITP) at the University of Colorado Boulder in 2017.

**Dr. Levi Perigo, University of Colorado Boulder**

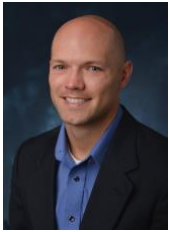

Dr. Perigo is a Scholar in Residence in Interdisciplinary Telecom Program (ITP) at the University of Colorado Boulder where he focuses on next generation networks, SDN/NFV, and network automation.